\documentclass{article}
\usepackage{spconf,amsmath,graphicx,subcaption,ulem}
\usepackage{xcolor}
\usepackage{booktabs}
\definecolor{myred}{RGB}{255, 45, 45}
\definecolor{myorange}{RGB}{251, 161, 63}
\definecolor{mygreen}{RGB}{70, 206, 115}
\definecolor{myblue}{RGB}{97, 142, 255}

\title{Multi-element microscope optimization by a learned sensing network with composite physical layers}
\name{Kanghyun Kim, Pavan Chandra Konda, Colin L. Cooke, Ron Appel, Roarke Horstmeyer}
\address{Pratt School of Engineering, Duke University, Durham, NC 27708, USA \thanks{Data and code: deepimaging.io}}

\begin{document}

%
\maketitle{}
\begin{abstract}
Standard microscopes offer a variety of settings to help improve the visibility of different specimens to the end microscope user. Increasingly, however, digital microscopes are used to capture images for automated interpretation by computer algorithms (e.g., for feature classification, detection or segmentation), often without any human involvement. In this work, we investigate an approach to jointly optimize multiple microscope settings, together with a classification network, for improved performance with such automated tasks. We explore the interplay between optimization of programmable illumination and pupil transmission, using experimentally imaged blood smears for automated malaria parasite detection, to show that multi-element "learned sensing" outperforms its single-element counterpart. While not necessarily ideal for human interpretation, the network’s resulting low-resolution microscope images (20X-comparable) offer a machine learning network sufficient contrast to match the classification performance of corresponding high-resolution imagery (100X-comparable), pointing a path towards accurate automation over large fields-of-view.
\end{abstract}

\section{Introduction}
The emergence of machine learning-based image analysis points to a new opportunity to re-examine microscope design. Unlike people, machine learning (ML) algorithms directly provide us with automated decisions that include quantitative measures of task-specific performance (e.g., the accuracy of feature classification or detection). These measures can subsequently allow us to objectively tune a microscope's design to improve the accuracy of automated ML outputs. Such a joint optimization approach, to design the microscope hardware for improved ML analysis, was recently demonstrated with a modified task-specific deep neural network termed a \textit{learned sensing network} (LSN) ~\cite{muthumbi2019learned}. In this work, we demonstrate a novel LSN that can jointly optimize a microscope's pupil transmission and specimen illumination for improved image classification. To enable end-to-end optimization of multiple microscope parameters within a task-specific neural network, we digitally model the physical transformations of the pupil and illumination within distinct \textit{physical layers}, prepended to our neural network. In this work, the physical layers are trained in conjunction with the neural network as a whole to optimize both the pupil transmission function and discrete illumination pattern, as detailed in Fig. \ref{fig:layers}.

Conceptually, pupil engineering and illumination control is not new. Computational imaging techniques \cite{mait2018computational} have been used for decades to extract new information by manipulating microscope parameters \cite{shechtman2016multicolour} or with radical new designs \cite{antipa2018diffusercam}. Recently, machine learning approaches are increasingly used to determine optimal microscope parameters \cite{sitzmann2018end, kellman2019physics}. However, these prior works mostly focus on improving the image quality, rather than the automation task performance \cite{hershko2019multicolor}.We have recently proposed task-guided optimization, but only for learning illumination patterns \cite{muthumbi2019learned, cooke2020physics}.

Here we aim to expand the role of machine learning in microscope design by optimizing multiple optical elements concurrently. By examining both the illumination and the pupil design, we can gain insights into the relative merits of applied coding schemes to each and attempt to understand their interplay. To aid our analysis, we considered two tasks: a simulated task using a synthetic dataset with known frequency characteristics, and an experimental task that explores multi-element optimization for detection of the malaria parasite within blood smears.

\section{Task specific deep optimization}
\begin{figure}[ht]
\centering
\includegraphics[width=1.0\linewidth]{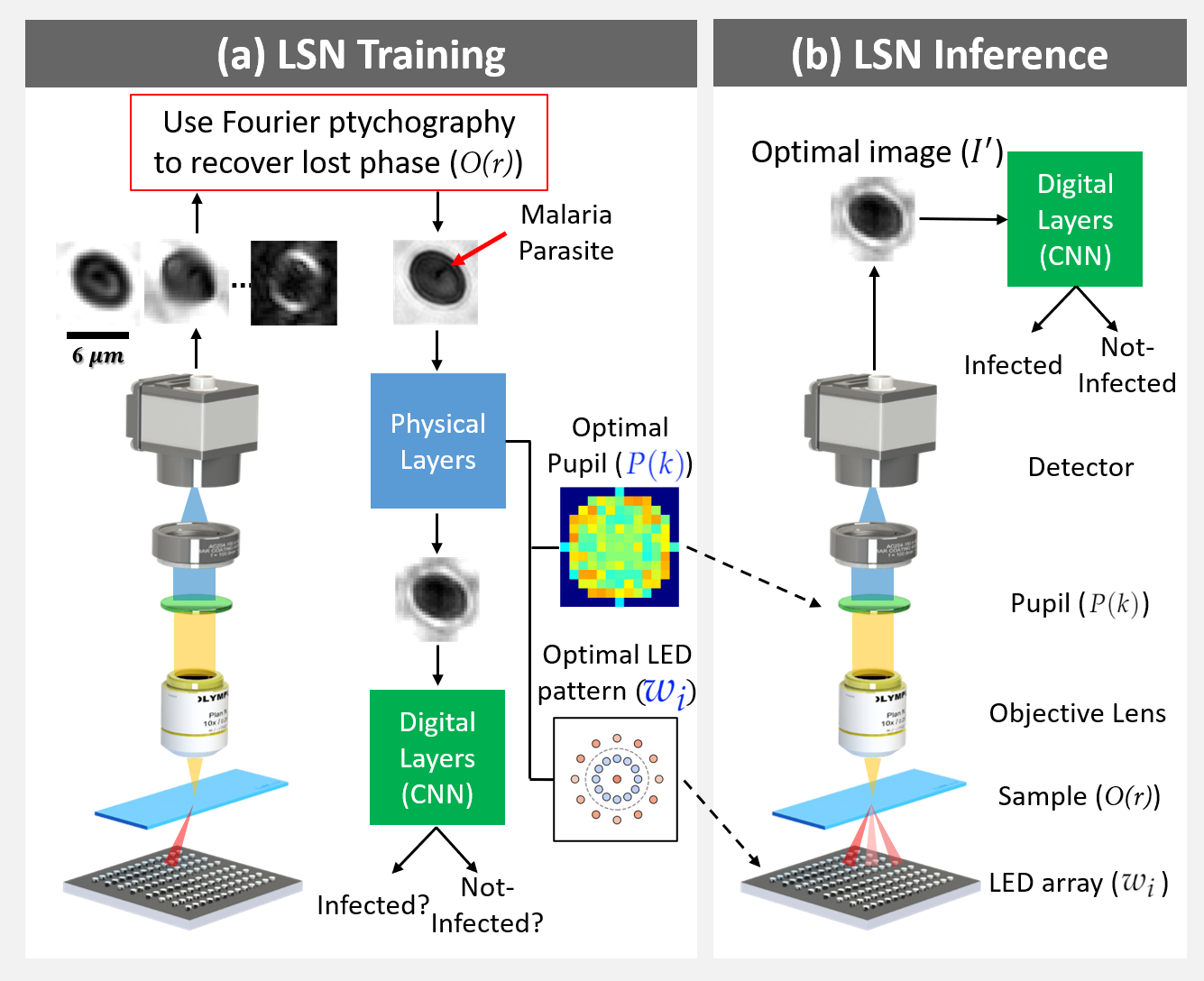}
\caption{Learned sensing network (LSN) framework: (a) During training, LEDs are turned on time-sequentially to capture multiple images, which are then processed with Fourier ptychography \cite{zheng2013wide, konda2020fourier} to recover the lost phase. Reconstructions are sent through the LSN physical layers (Eq. \ref{eq:simulation_equation}), facilitating the optimization of pupil transmission and illumination weights, culminating in an optimized image for LSN digital layer processing. Supervised learning is used to optimize both physical and digital layer weights. (b) During inference, the optimized pupil transmission and LED illumination pattern are used to generate the optimal image at the detector, which directly enters the digital CNN layers for accurate classification.
}
\label{fig:layers}
\end{figure}

Prepending physical layers to a deep convolutional neural network (CNN) is key to our LSN. Physical layers provide a data-informed forward model for image formation that includes the optical elements that we aim to optimize. We start our microscope's forward model by assuming that we have a thin sample $O(r)$, which is illuminated with a plane wave traveling with transverse wave vector $k_i$ generated by a quasi-monochromatic LED placed beneath the specimen (Fig. \ref{fig:layers}). This LED is one of many within an array of $i=1$ to $n$ LEDs, each generating plane waves at slightly different angles $k_i$ for selective sampling of the sample's spatial frequencies \cite{zheng2013wide}. Illumination from a single LED will result in a shifted spectrum $\hat{O}(k-k_i)$ at the microscope back focal plane. This spectrum is filtered by the pupil transmission function $P(k)$ (amplitude only) before propagating to the detector plane. With multiple ($i=1:n$) mutually incoherent LEDs illuminated simultaneously, multiple shifted spectra are filtered before adding incoherently at the detector plane. The resulting image formed by $n$ LEDs turned on at different normalized brightnesses ($w_i$) can be generated via the weighted intensity sum \cite{muthumbi2019learned}:

\begin{equation}
 I' = \sum_{i=1}^n \{ |\mathcal{F}^{-1}[\hat{O}(k - k_i) \times \textcolor{blue}{P(k)}]|^2 \times \textcolor{blue}{w_i} \}
\label{eq:simulation_equation}
\end{equation}

This equation is our physical layer's forward model. In this work, we aim to optimize for both the LED brightnesses $w_i$ and the pupil function $P(k)$, which form our composite physical layers. As is clear in Eq. \ref{eq:simulation_equation}, this joint optimization task is nonlinear and thus often does not permit straightforward solutions. After detection, our goal is to classify the resulting image $I'$ with a deep neural network, which forms the \textit{digital} portion of our LSN (i.e., its \textit{digital layers}). During training, we minimize a loss based on the classification output of the LSN and backpropagate the intermediate errors to optimize parameters in both our physical and digital layers. During inference, we use the optimized parameterization of our physical layers to configure the experimental setup, and use our digital layers to classify the resulting image $I'$.

To understand the effectiveness of joint optimization on the pupil transmission and the illumination weights, and to study the interplay between them, we consider the following four training cases:

\begin{enumerate}
    \item \textbf{Digital-only Optimization (DO):} Illumination and pupil transmission are fixed to default values - only the center LED is illuminated (non-zero) and all pupil weights set to one (i.e., a clear aperture).
    \item \textbf{Pupil Optimization (PO):} The pupil transmission is optimized, while the illumination is fixed to the default value of normalized plane-wave illumination.
    \item \textbf{Illumination Optimization (IO):} Illumination weights are optimized, while the pupil is fixed to its default value.
    \item \textbf{Pupil and Illumination Optimization (PIO):} Both pupil transmission and illumination weights are optimized.
\end{enumerate}

The network architecture remains identical in all four cases but the physical layers' weights are updated (or not) according to each case. For the LSN digital layers, we used four convolution layers (six channels per layer with max pooling after the second and fourth layers), followed by two fully-connected layers (with 64 and 2 hidden units respectively). A ReLU activation was used after each layer except the final one (softmax function).

\section{Simulations}
\begin{figure}[htb]
  \centering
  \centerline{\includegraphics[width=1.0\linewidth]{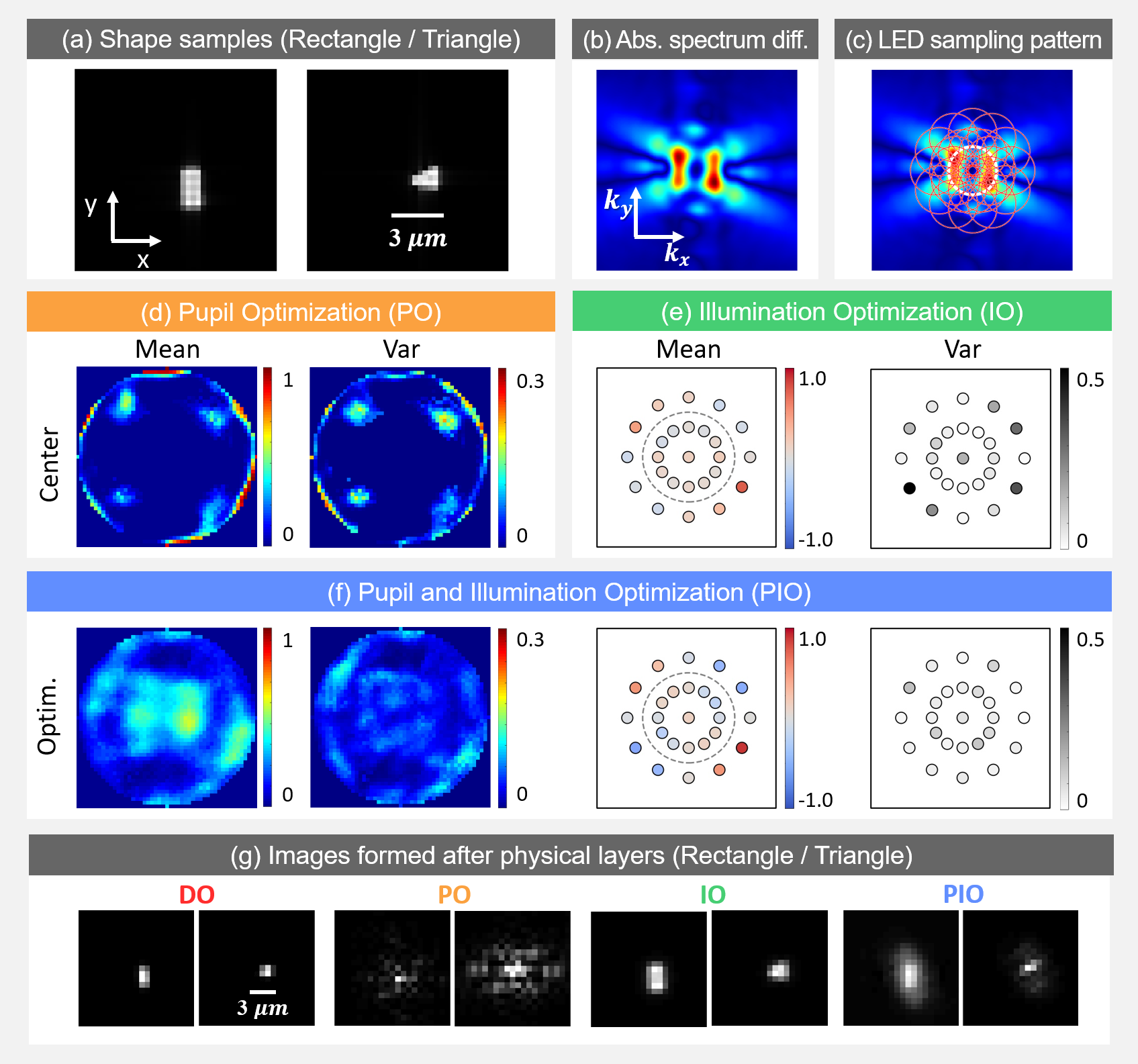}}
\caption{Simulation results. (a) Example rectangular and triangular samples (absorption map). (b) Absolute value of average Fourier spectra difference between two sample types. (c) Marked circles represent spatial frequencies sampled by each corresponding LED illumination angle. (d) Mean and variance of pupil optimization results. Trained pupil is thresholded for better contrast before plotting. (e) Illumination optimization results. (f) Pupil and Illumination optimization results. (g) Example images.}
\label{fig:simulation_data_results}
\end{figure}

To better understand the interplay between illumination and pupil transmission optimization, we first designed a classification task based on a synthetic dataset to differentiate between hand-drawn triangles and rectangles. As these two shapes have distinct spatial frequency characteristics (shown in Fig. \ref{fig:simulation_data_results} (b)), we hypothesized that the optimized pupil and illumination patterns would selectively highlight the spatial frequency regions in which the average difference in the frequency spectra of the two shapes were maximally different, to subsequently improve classification performance.

The synthetic dataset consisted of 300 examples of each shape, drawn on an \textit{iPad} and exported as $256\times256$ pixel images. These images were then zero-padded to $2048\times2048$ pixels and augmented eight times by translation, resulting in a total of 2700 samples per shape. We defined our complex-valued thin sample as $O(x,y) = A(x,y) \times exp($i $\phi(x,y))$, where $A(x,y)$ is a normalized hand-drawn image and $\phi(x,y)$ is the phase transmittance of the sample. Here for simplicity, we set $\phi(x,y)$ as $2A(x,y)$. We then simulated microscope image formation following Eq. \ref{eq:simulation_equation}, assuming a 10X objective lens ($0.2$ NA, 522 nm illumination wavelength), with a variable-transmission pupil function (49 pixels diameter), and a $64\times64$ pixel sensor. The simulated illumination array included $n=25$ LEDs in three concentric rings, illuminating at $0^\circ$, $16.37^\circ$, and $34.30^\circ$ angles, such that the 13 innermost LEDs were bright-field and the 12 outermost were dark-field. The frequencies sampled by these different illumination angles are marked in Fig. \ref{fig:simulation_data_results} (c)). $1\%$ Gaussian detector noise was added to each simulated image and both positive and negative values are allowed for LED weights, which translates to capturing two images (one with positive weights, another with negative weights) and subtracting them for the resultant image $I'$ in Eq. \ref{eq:simulation_equation}.


The LSN was optimized for four training cases discussed above and each case was run 15 times with different random initialization seeds. The classification results from these configurations are shown in Table \ref{tab:table_simulation_data_results}, with the best performing results in bold. We observe that joint optimization (\textbf{PIO}) yields the highest performance, with an average classification accuracy of $99\%$, while the digital-only optimization (\textbf{DO}) has the lowest overall performance ($80\%$). The cases where only a single element is optimized (\textbf{PO} and \textbf{IO}) offers significant improvement from the \textbf{DO} case ($90\%$ and $92\%$ respectively) although still lower than the joint optimization case. The mean and variance of the normalized optimized patterns (pupil transmissions and illumination weights) across multiple runs are shown in Fig. \ref{fig:simulation_data_results}.


For the \textbf{PO} case, the optimized pupil transmission function selects spatial frequencies concentrated in four areas (Fig. \ref{fig:simulation_data_results} (d)), which are consistent with areas highlighted by the average difference between the frequency spectra (center white ring in Figure \ref{fig:simulation_data_results} (c)). By selectively transmitting these "difference" spatial frequencies, and attenuating others, the resulting classifier better distinguishes between the two shape categories.

When optimizing only the sample illumination (\textbf{IO}), the converged LED pattern (shown in Fig. \ref{fig:simulation_data_results} (e)) primarily emphasized two dark-field LEDs along the anti-diagonal. Within our illumination optimization, we generally observed a tendency to emphasize dark-field LEDs, since they help transmit higher spatial frequencies to the image sensor. In this case since the average Fourier spectra difference shown in Fig. \ref{fig:simulation_data_results} (b) is larger in the anti-diagonal regions, we hypothesize that the network converged upon this dark-field LED grouping to selectively highlight these differences, similar to the pupil transmission function.


The joint pupil and illumination optimization case (\textbf{PIO}, Fig. \ref{fig:simulation_data_results} (f)) yielded the most interesting results. Here, we observed how the two components worked together to produce a microscope design for the most accurate sample classification. The converged \textbf{PIO} pupil was distinct from the \textbf{PO} pupil and the optimized illumination pattern one again exhibits more energy on the diagonal and anti-diagonal regions using positive and negative LED weights, but in a unique pattern. One direct observation regarding the interplay between the two coding elements is that in the joint-optimization case, the optimized pupil offers DC transmission, while in the pupil-only case it does not. This is likely due to the ability to utilize the LED illumination to provide primarily dark-field illumination. A second observation is that in the joint optimization case, the resulting images in Fig. \ref{fig:simulation_data_results} (g) are spread over fewer pixels than in the pupil-only case, leading to higher SNR in the presence of sensor noise. Finally, comparing the normalized optical transmission of each case yields an interesting insight - the \textbf{PO} mask transmits 5\% of incident light, the \textbf{IO} case emits 14\% of the maximum normalized light from the LED array, and the total emission and transmission of the \textbf{PIO} is also 5\% - suggesting an inherent trade-off between illumination and pupil coding versus light transmission in learned sensing design.

\renewcommand{\arraystretch}{1.1}
\begin{table}[htb]
  \caption{\bf LSN classification accuracy, triangles and rectangles}
  \label{tab:table_simulation_data_results}
  \centering
  \begin{tabular}{cccc}
    \toprule
    \textcolor{myred}{Digital(DO)} & \textcolor{myorange}{Pupil(PO)} & \textcolor{mygreen}{Illu.(IO)} & \textcolor{myblue}{Joint(PIO)} \\
    \midrule
    79.5 $\pm$ 10.6 & 90.2 $\pm$ 6.0 & 92.1 $\pm$ 8.1 & \textbf{99.0 $\pm$ 1.5} \\
    \bottomrule
  \end{tabular}
  \hspace{0.1cm}
\end{table}

\section{Experiments}
To demonstrate our pupil and illumination LSN optimization with real-world data, we turned to the task of detecting malaria-infected blood cells in a thin blood smear. We used the malaria infected blood cell images dataset reported in \cite{muthumbi2019learned, ou2015high}, which was captured on a microscope using a 20X objective lens under 29 illumination angles from an LED array. The dataset consisted of 328 infected and 693 uninfected cropped blood cells, where each uniquely illuminated image for each cell is $28\times28$ pixels. Since our model in Eq. \ref{eq:simulation_equation} requires us to know the complex thin sample transmission function for each cell (i.e., its absorption and phase), we used Fourier ptychography \cite{konda2020fourier} to recover the unknown phase from the 29 low-resolution intensity images, which resulted in high-resolution complex-valued reconstructions with $112\times112$ pixels. These high-resolution images were augmented three times by rotations to generate a final dataset of 1312 infected and 2772 uninfected complex thin sample transmission functions $O(r)$, which we input into the LSN physical layer in Eq. \ref{eq:simulation_equation} to produce a $28\times28$ pixel image, which then entered the LSN digital layers for binary classification (infected vs. not-infected). The modeled microscope included a 20X objective lens with $0.4$ NA, $5.5 \mu m$ detector pixel size and green illumination with 522 nm wavelength. The modeled illumination for the LSN physical layer arose from a  set of 25 LEDs that differed in position from the original 29 LEDs to avoid systematic errors during network training (3 concentric circles at angles $0^\circ$, $19.81^\circ$, and $48.22^\circ$, 13 bright-field and 12 dark-field).

We tested the same four training cases outlined above, repeating each 50 times under random initializations. In Table \ref{tab:table_experiment_data_results} sensitivity and specificity are reported in addition to the accuracy due to the uneven split of data in the two classes. We also performed classification directly on the high-resolution $112\times112$ images for a reference point. We observe that once again the \textbf{PIO} configuration, which jointly optimizes both the illumination and pupil, yields the best results with performance comparable to using the high-resolution images. However, as shown by the performance of the \textbf{IO} configuration, the majority of the gains appear to be due to illumination optimization, while the pupil only case (\textbf{PO}) offers a marginal improvement from baseline (90.0\% vs. 89.7\% accuracy). The performance of the combined pupil and illumination optimization (\textbf{PIO}) shows that by collectively optimizing these elements we can achieve higher performance than by optimizing a single coding element.

\begin{figure}[htb]
  \centering
  \centerline{\includegraphics[width=1.0\linewidth]{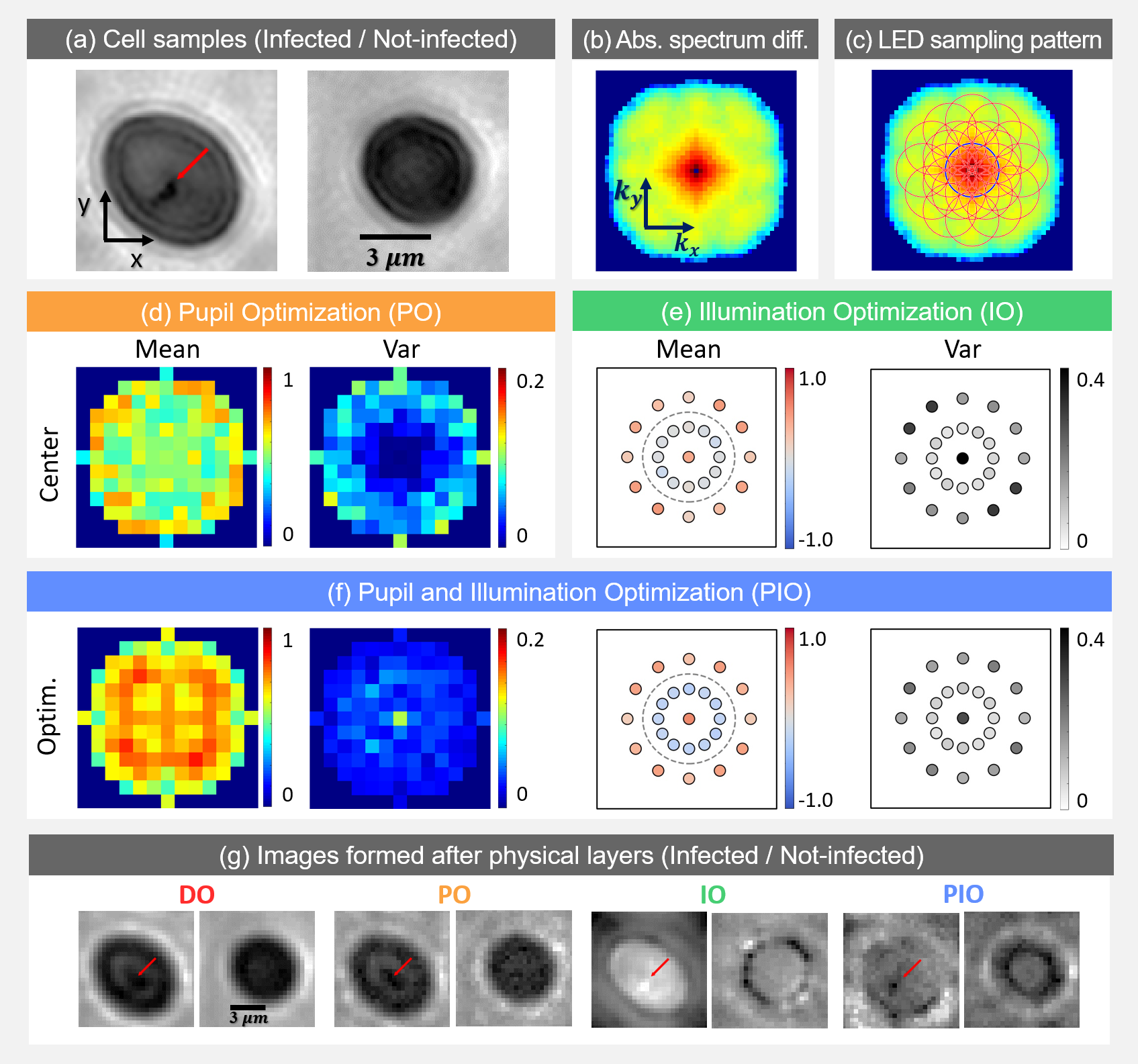}}
\caption{Malaria parasite detection results. (a) Sample high-resolution images of infected (see below) and not-infected cells. (b) Log plot of absolute value of Fourier spectra difference between two cell categories. (c) Each red circle represents spatial frequencies sampled by the corresponding illumination angle. (d) Mean and variance of pupil optimization results. (e) Illumination optimization results. (f) Pupil and Illumination optimization results. (g) Example images.}
\label{fig:experiment_data_results}
\end{figure}

The mean and variance of the optimized pupils and LED weights are plotted in Fig. \ref{fig:experiment_data_results}. We also show the difference of the average spectra of the two sample categories in Fig. \ref{fig:experiment_data_results} (b), but unlike in our simulation, this difference also contains information about red blood cell shape, background variation, noise and reconstruction artifacts. Hence, it is more challenging to gain direct insight into the optimized coding elements with this experimental result. However, we know a priori that the parasites are approximately 1 $\mu$m in size, so the primary differences of interest surround this higher spatial frequency, and we can expect an optimized system to preferentially reduce the DC component and increase these higher frequencies with dark-field illumination. This trend was observed in the illumination patterns for both \textbf{IO} and \textbf{PIO} cases, which resulted in relatively sharp dark-field-type images as shown in Fig. \ref{fig:experiment_data_results} (e/f), where the parasite location tended to appear as a highlighted spot with different contrast with respect to the background (Fig. \ref{fig:experiment_data_results} (g)).

\renewcommand{\arraystretch}{1.1}
\begin{table}[htb]
  \caption{\bf LSN classification accuracy, malaria-infected blood}
  \label{tab:table_experiment_data_results}
  \centering
  \begin{tabular}{cccc}
    \toprule
    Case & Acc. & Sens. & Spec.   \\
    \midrule
    \textcolor{myred}{Digital(DO)} & 89.7 $\pm$ 2.6 & 72.2 $\pm$ 8.0 & 98.4 $\pm$ 0.7\\
    \midrule
    \textcolor{myorange}{Pupil(PO)} & 90.0 $\pm$ 2.1 & 74.9 $\pm$ 6.8 & 97.5 $\pm$ 1.1\\
    \midrule
    \textcolor{mygreen}{Illu.(IO)} & 97.3 $\pm$ 1.9 & 95.8 $\pm$ 4.2 & 98.0 $\pm$ 2.1 \\
    \midrule
    \textcolor{myblue}{Joint(PIO)} & \textbf{98.8 $\pm$ 1.5} & \textbf{97.7 $\pm$ 3.6} & 99.3 $\pm$ 1.0 \\
    \midrule
    \textcolor{black}{High-NA} & 97.0 $\pm$ 0.7 & 92.2 $\pm$ 2.0 & \textbf{99.4 $\pm$ 0.5} \\
    \bottomrule
  \end{tabular}
\end{table}

Examining the optimized pupil and illumination patterns themselves, we observe a similar trend as in the simulation, where the \textbf{PO} case attenuates the DC component and selectively transmits higher spatial frequencies. However, the low NA objective lens does not transmit the parasite frequencies at high fidelity, hence \textbf{PO} optimization doesn't have a significant impact on the results. In the joint \textbf{PIO} case, it is once again difficult to interpret the pupil shape, but it appears to follow a trend similar to the pupil in our simulation. Here, the illumination optimization is now preferentially passing higher spatial frequencies corresponding to parasite features in both the \textbf{IO} and \textbf{PIO} cases. From these results, it is clear that illumination optimization provides a significant performance boost. However, the additional degrees of freedom in the multi-element optimization case has the best overall performance.


\section{Discussion and future work}
In this work, we demonstrated the ability to simultaneously optimize multiple optical coding elements to improve the automated classification performance of a modified deep CNN. Our learned sensing approach for multiple elements provides better performance than optimizing any individual optical element alone. Here, the optical parameters are optimized for a specific task and sample type, which can be practically implemented using programmable optical elements, such as the LED array used here or with a digital micro-mirror device (DMD), to produce an intelligent imaging system that can quickly change its settings based on the task at hand. 

We only considered amplitude pupil transmission in our first demonstration. In future work, pupil phase can also be considered within a higher optimization space. Other optical parameters and coding elements can also be easily optimized within our proposed framework. Illumination wavelength and polarization can be directly added in, as can alternative optical coding elements at planes besides the back focal plane.

The training of our forward model required knowledge of specimen phase, which we estimated with Fourier ptychography. Alternate approaches for phase recovery, such as digital holography or integrating the Fourier ptychographic reconstruction within a neural network \cite{konda2020fourier}, are interesting future avenues to pursue. Finally, we have chosen a classification task for this work, however, this joint-optimization approach can be easily expanded to other machine learning tasks such as image segmentation or virtual fluorescence microscopy \cite{cooke2020physics}.



\noindent\textbf{Acknowledgments.} Authors would like to thank Xiang Dai for help with Fourier ptychographic reconstruction and Kevin Zhou for feedback on the manuscript.

\newpage
\bibliographystyle{IEEEbib}
\bibliography{refs}

\end{document}